\documentclass[apj]{emulateapj}

\usepackage{amsmath}
\usepackage{empheq}

\shortauthors{LAPI \& DANESE}
\shorttitle{A STOCHASTIC THEORY OF HIERARCHICAL CLUSTERING I.}
\slugcomment{ACCEPTED BY APJ}

\newcommand*\widefbox[1]{\fbox{\hspace{2em}#1\hspace{14.5em}}}

\begin{document}

\title{A Stochastic Theory of the Hierarchical Clustering\\
\small{I. Halo Mass Function}}

\author{Andrea Lapi\altaffilmark{1,2,3,4}, Luigi Danese\altaffilmark{1,2}}\altaffiltext{1}{SISSA, Via Bonomea 265, 34136 Trieste, Italy}\altaffiltext{2}{IFPU - Institute for fundamental physics of the Universe, Via Beirut 2, 34014 Trieste, Italy}\altaffiltext{3}{INFN-Sezione di Trieste, via Valerio 2, 34127 Trieste,  Italy}\altaffiltext{4}{INAF-Osservatorio Astronomico di Trieste, via Tiepolo 11, 34131 Trieste, Italy}

\begin{abstract}
We present a new theory for the hierarchical clustering of dark matter (DM) halos based on stochastic differential equations, that constitutes a change of perspective with respect to existing frameworks (e.g., the excursion set approach); this work is specifically focused on the halo mass function. First, we present a stochastic differential equation that describes fluctuations in the mass growth of DM halos, as driven by a multiplicative white (Gaussian) noise dependent on the spherical collapse threshold and on the power spectrum of DM perturbations. We demonstrate that such a noise yields an average drift of the halo population toward larger masses, that quantitatively renders the standard hierarchical clustering. Then, we solve the Fokker-Planck equation associated to the stochastic dynamics, and obtain the Press \& Schechter mass function as a (stationary) solution. Moreover, generalizing our treatment to a mass-dependent collapse threshold, we obtain an exact analytic solution  capable of fitting remarkably well the $N$-body mass function over a wide range in mass and redshift. All in all, the new perspective offered by the theory presented here can contribute to better understand the gravitational dynamics leading to the formation, evolution and statistics of DM halos across cosmic times.
\end{abstract}


\section{Introduction}\label{sec|intro}

The halo mass function, namely the statistics describing the number of dark matter (DM) halos per unit comoving volume as a function of halo mass and redshift, is a fundamental quantity in astrophysics and cosmology (see textbooks by Mo et al. 2010 and Cimatti et al. 2020). For example, it is a basic ingredient to develop sensible galaxy formation and evolution models (see reviews by Silk \& Mamon 2012 and Naab \& Ostriker 2017), and it is routinely exploited in cosmological studies relying on the abundance and clustering of collapsed objects and of large-scale structures (see reviews by Frenk \& White 2012 and Wechsler \& Tinker 2018).

Clearly, the halo mass function can be estimated via high-resolution, large-volume, cosmological $N$-body simulations (see Sheth \& Tormen 1999; Jenkins et al. 2001; Warren et al. 2006; Tinker et al. 2008; Crocce et al. 2010; Bhattacharya et al. 2011; Watson et al. 2013). However, given the natural limits on resolution, computational time, and storing capacity, it can be probed only in limited mass and redshift ranges. Moreover, the results of simulations depend somewhat on the algorithm used to identify collapsed halos (e.g., friend-of-friend vs. spherical overdensity), and on specific parameters related to the identification of isolated objects (e.g., the linking length). On the other hand, to estimate the halo mass function from observations is even more challenging (e.g., Castro et al. 2016; Dong et al. 2019; Sonnenfeld et al. 2019; Li et al. 2020), given the statistical and systematic uncertainties that arise when linking the observable quantities to the halo mass. Therefore, a deep theoretical understanding on how the halo mass function is originated from first principles is of crucial importance.

The modern theoretical framework to address the issue was born with the seminal work by Press \& Schechter (1974); these authors were able to compute an analytic expression for the halo mass function by prescribing that a halo would collapse if it resided within a sufficiently overdense region of the initial (Gaussian) perturbation field. Given that the overdensity around a spatial location depends on scale, they recognized that the halo abundance is simply related to the mass fraction in the density field, smoothed on different scales, which is above a critical threshold for collapse. A drawback of this approach is the so called 'cloud-in-cloud' problem, i.e.,  attention must be paid not to double count overdense regions embedded within a larger collapsing perturbation; in other words, one has to consider only the perturbations that overcome the threshold on a given smoothing scale, but not on a larger one.

The problem was solved with the development of the excursion set framework by  Bond et al. (1991), which still nowadays constitutes the standard theory. This envisages the overdensity around a given spatial location to execute a random walk when considered as a function of the smoothing scale; if the smoothing is performed with a sharp filter in Fourier space, the walk is Markovian.  The collapse threshold here plays the role of a barrier, and the halo mass function is related to the distribution of first crossing, i.e., the probability that a walk crosses the barrier for the first time on a specific scale. In the original theory, the collapse threshold was gauged on the spherical collapse model of DM perturbations (Gunn \& Gott 1972), and as such it was assumed to be independent on halo mass; subsequent developments adopted a mass-dependent threshold, inspired from the ellipsoidal collapse model (see Sheth \& Tormen 2002), in order to better reproduce the results of $N$-body simulations.

The excursion sets approach was then exploited to derive the 'conditional' halo mass function (see Lacey \& Cole 1993), describing the mass and redshift distribution of a halo's progenitors, to build up Monte Carlo realizations of the merging process known as merger trees (see Kauffmann \& White 1993; Somerville \& Kolatt 1999; Cole et al. 2000; Parkinson et al. 2008), and to develop models for the large-scale halo bias (see Mo \& White 1996; Sheth \& Lemson 1999). Further, more recent, refinements include non-Markovian walks (Maggiore \& Riotto 2010a; Musso \& Sheth 2012), stochastic collapse thresholds (Maggiore \& Riotto 2010b; Corasaniti \& Achitouv 2011), extension to peaks theory (see Paranjape et al. 2012), descriptions of the void distribution (see Sheth \& van de Weygaert 2004; Jennings et al. 2013), and non-standard cosmologies (e.g., von Braun-Bates \& Devriendt 2018; Lovell 2020).

Despite this rich literature focused on the theoretical foundations and a number of undoubtable successes in practical applications, the excursion set framework is known to hide some pitfalls and drawbacks: no exact analytic expression of the mass function for a general mass-dependent collapse threshold is known, but for very simple shapes (see Zhang \& Hui 2006; Lapi et al. 2013); the merging kernel associated to the excursion set theory is not symmetric, and this causes a mathematical inconsistency or at least an ambiguity in the definition of the merger rates (see Benson et al. 2005; Neistein \& Dekel 2008; Zhang et al. 2008); even adopting a mass-dependent collapse threshold and other refinements, the excursion set formalism struggles to reproduce the halo progenitors' distributions extracted from $N$-body simulations (e.g., Parkinson et al. 2008; Jiang \& van den Bosch 2014); the relation between the probability of first upcrossing and the mass function, which is at the heart of the excursion set framework, is correct on statistical grounds but cannot be strictly true for individual mass elements (see discussion by Mo et al. 2010, their Sect. 7.2.2b).

In this paper we submit a new theory of the hierarchical clustering and halo mass function based on stochastic differential equations in real space, that constitutes a change of perspective with respect to the excursion set formalism. First, we invent a stochastic differential equation that describes fluctuations in the mass growth of DM halos, as driven by a multiplicative white (Gaussian) noise dependent on the spherical collapse threshold and on the power spectrum of DM perturbations; in this approach it is the mass (or mass variance) in a given region of the Universe to perform a (Markovian) random walk as a function of cosmic time. In Sect.~\ref{sec|whitenoise} we demonstrate that the noise yields an average drift toward larger masses, that quantitatively renders the standard hierarchical clustering. Then, in Sect.~\ref{sec|fokker}  we solve the Fokker-Planck equation associated to the stochastic dynamics, and obtain as a solution the Press \& Schechter mass function;  in Sect.~\ref{sec|stationary} we point out that the solution is stationary when the original equation is written in convenient variables.

In Sect.~\ref{sec|barrier} we introduce a minimal modification of the stochastic equation in terms of a mass-dependent collapse threshold. Using a parametric shape analogous to that adopted in the excursion set framework, we obtain a closed-form analytical solution of the associated Fokker-Planck equation. Remarkably, such a solution has a shape similar to the empirical fitting formula introduced since Sheth \& Tormen (1999); moreover, for specific values of the parameters describing the mass dependence of the collapse threshold, our result reproduces remarkably well the $N$-body mass function over an extended range of masses and redshifts.

As an aside, in Sect.~\ref{sec|colnoise} we explore how to generalize our framework when a colored rather than a white noise is considered, so as to enforce a non-Markovian evolution. Adopting for definiteness a multiplicative Ornstein-Uhlenbeck noise and a constant threshold for collapse, we are able to solve exactly the corresponding Fokker-Planck equation and obtain a closed form solution. With respect to the white noise case, the redshift evolution of the mass function is found to be modified somewhat, in a fashion depending on the correlation time characterizing the noise. Finally, in Sect.~\ref{sec|summary} we summarize our findings and envisage possible outlooks.

Throughout this work, we adopt the standard flat $\Lambda$CDM cosmology (Planck Collaboration 2018) with rounded parameter values: matter density $\Omega_M\approx 0.3$, dark energy density $\Omega_\Lambda\approx 0.7$, baryon density $\Omega_{\rm b}\approx 0.05$, Hubble constant $H_0 = 100\,h$ km s$^{-1}$ Mpc$^{-1}$ with $h\approx 0.7$, and mass variance $\sigma_8\approx 0.8$ on a scale of $8\, h^{-1}$ Mpc. The
most relevant expressions are highlighted with a box.

\section{A stochastic equation for the hierarchical clustering}\label{sec|whitenoise}

Our proposal is to capture the essence of the hierarchical clustering via the following nonlinear stochastic differential equation:
\begin{empheq}[box=\fbox]{align}\label{eq|basic_sigma}
\cfrac{\rm d}{{\rm d}t}\ln\sigma^{-1} = \cfrac{\sigma}{\delta_c}\, \left|\cfrac{\dot\delta_c}{\delta_c}\right|^{1/2}\, \eta(t)~,
\end{empheq}
or equivalently, in terms of mass\footnote{Throughout the paper we adopt, in line with the majority of the physics community, the Stratonovich convention; this allows to use the rules of ordinary calculus on stochastic variables, at the price of originating a noise-induced drift term in the Fokker-Planck coefficients associated to a given stochastic differential equation (see Appendix A). The alternative Ito convention, mostly used by mathematicians, removes such a noise-induced drift, but requires to develop new rules for the differential calculus of stochastic variables.}
\begin{equation}\label{eq|basic_mass}
\cfrac{\rm d}{{\rm d}t}\, M = \cfrac{\sigma^2}{|{\rm d}\sigma/{\rm d}M|}\, \cfrac{1}{\delta_c}\, \left|\cfrac{\dot\delta_c}{\delta_c}\right|^{1/2}\, \eta(t)~.
\end{equation}
Here $\eta(t)$ is a Gaussian white noise (physical dimension $1/\sqrt{\rm time}$) with ensemble-average properties $\langle\eta(t)\rangle=0$ and $\langle\eta(t)\, \eta(t')\rangle=2\, \delta_{\rm D}(t-t')$, where $\delta_{\rm D}$ is the Dirac delta-function (the factor $2$ is only a convention and clearly it could be reabsorbed into the multiplicative term). The quantity $\delta_c(z)=\delta_{c0}\,D(0)/D(z)$ is the critical threshold for collapse extrapolated from linear perturbation theory; in a flat Universe (see Mo et al. 2010; Weinberg 2008), one can use the approximations $\delta_{c0} \simeq \frac{3}{20}\, (12\pi)^{2/3}$ $\left[1+0.0123\log_{10}\Omega_M(z)\right]\approx 1.68$
and $D(z) \approx \frac{5}{2}\,\frac{\Omega_{M}(z)}{1+z}\,\left[\frac{1}{70}+
\frac{209}{140}{\Omega_M(z)-\frac{1}{140}\,\Omega_M^2(z)+\Omega_M^{4/7}(z)}\right]^{-1}$ with $\Omega_M(z)\equiv \Omega_M\,(1+z)^3/[\Omega_\Lambda+\Omega_M\,(1+z)^3]$.
Finally, $\sigma$ is the mass variance filtered on the mass scale $M$:
\begin{equation}\label{eq|variance}
\sigma^2(M) = \cfrac{1}{(2\pi)^3}\,\int{\rm d}^3k\, P(k)\,\tilde{W}_M^2(k)~;
\end{equation}
here $P(k)$ is the power spectrum of density fluctuation, and $\tilde{W}_M^2(k)$ is the Fourier transform of a window function whose volume in real space encloses the mass $M$; for standard cold dark matter power spectra (e.g., Bardeen et al. 1986), $\sigma(M)$ is an inverse, convex, slowly-varying function of $M$. Note that in the present theory the relation $\sigma(M)$ between $\sigma$ and $M$ is purely deterministic, but both $M(t)$ and $\sigma(t)=\sigma(M(t))$ are to be considered stochastic variables that fluctuate over cosmic time $t$ under the influence of the noise. We stress the change of perspective with respect to the standard excursion set formalism: in the latter the overdensity field $\delta(\sigma)$ executes a random walk as a function of the mass variance $\sigma$, which plays the role of a pseudo-time variable; here the mass $M(t)$ or the mass variance $\sigma(M(t))$ are themselves stochastic variables, undergoing a Markovian evolution as a function of (real) time $t$.
Note that in the excursion set approach the choice of the filter function in Eq.~(\ref{eq|variance}) has a crucial impact, since the random trajectories $\delta(\sigma)$ are Markovian only when a sharp filter in
Fourier space is adopted; in the present theory, assuming a different filter
function (e.g., Gaussian or top-hat in real space) changes only the deterministic relation $\sigma(M)$ but has otherwise no effect on the Markovianity of the stochastic processes $M(t)$ and $\sigma(M(t))$.

\begin{figure}[!t]
\centering
\includegraphics[width=\columnwidth]{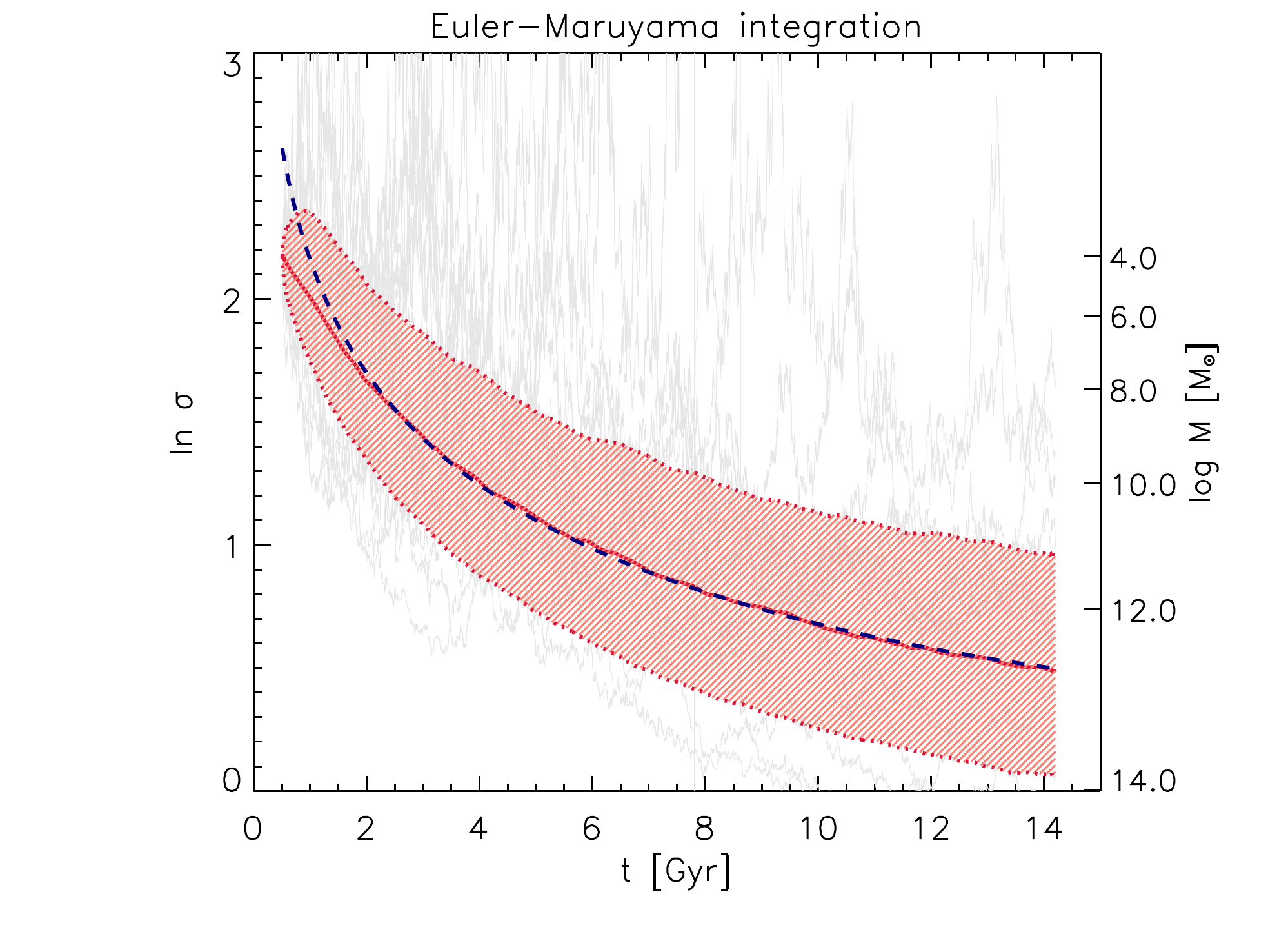}
\caption{Euler-Maruyama integration of the stochastic differential Eq.~(\ref{eq|basic_sigma}), yielding
the evolution of the mass variance $\ln\sigma$ (left $y-$axis) and of the mass $M$ (right $y-$axis) as a function of cosmic time $t$; the initial condition $M\approx 10^4\, M_\odot$ at $z\sim 10$ (corresponding to $t\sim 0.5$ Gyr) has been adopted. The grey lines are $30$ randomly chosen evolutionary tracks, while the red line and shaded area illustrate the median and the quartiles over $3000$ realization of the noise; the dashed blue line shows the evolution of the spherical collapse threshold $\delta_c(t)$.}\label{fig|sims}
\end{figure}

The rationale naively followed to invent the above Eq.~(\ref{eq|basic_sigma}) is simple. On the left hand side it appears the time derivative of an adimensional function of the mass that incorporates the power spectrum and the filtering scale; in choosing $\ln\sigma$ we have been inspired by a number of $N$-body simulations (e.g., Zhao et al. 2009), that suggest the mass growth of halos to be easily described in terms of such a quantity. On the right hand side it appears a stochastic driving $\eta(t)$, that for dimensional consistency must be multiplied by the (inverse) square root of a characteristic timescale. Since our aim here is to describe the growth of DM perturbations using quantities related to the linear regime, we find it natural to choose the timescale $|\dot \delta_c/\delta_c|\simeq |\dot D(t)/D(t)|$, where the factor $D(t)$ defined below Eq.~(\ref{eq|basic_mass}) effectively describes the linear growth of perturbations under gravity in a given cosmological background. Moreover, the adopted noise is `multiplicative', in that its strength depends on the state of the system, and specifically on the ratio $\sigma/\delta_c$. Regions with $\sigma\gtrsim \delta_c$ tend to change their mass more abruptly, while the evolution is slower for $\sigma\lesssim \delta_c$. In particular, positive variations of $M$ (or in $\ln\sigma^{-1}$) within the filtered region can be reasonably related to mergers among collapsed halos, or mass accretion from the field; negative variations can be interpreted in terms of mass loss due to gravitational interactions with surrounding regions, tidal forces, stripping, and fragmentation. We stress the crucial role played by the multiplicative noise in Eq.~(\ref{eq|basic_sigma}); as $\eta(t)$ fluctuates, also the random variable $\sigma$ and hence the multiplicative factor $\sigma/\delta_c$ on the r.h.s. varies, and therefore $\langle\sigma\, \eta/\delta_c\rangle$ is not null even if $\langle\eta\rangle$ is;
this noise-induced drift actually makes $\sigma(t)$ to copy the decrease of $\delta_c(t)$ with time and, given the inverse convex shape of the deterministic function $\sigma(M)$, a net average increase in mass $M(t)$ is enforced. Similar stochastic models with multiplicative noise have been employed to describe a wide range of physical phenomena, from Brownian motion in inhomogeneous media or in close approach to physical barriers, to thermal fluctuations in electronic circuits, to the evolution of stock prices, to the heterogeneous response of biological systems and randomness in gene expression; to our knowledge, this is the first time they are applied to describe the formation of collapsed structure in the Universe.

A more quantitative hint that such a stochastic equation has some value can be derived by performing a simple integration via the Euler-Maruyama method (e.g. Kloeden \& Platen 1992; there are other ways to obtain more accurate numerical solutions of stochastic differential equations but these are not needed here); Eq.~(\ref{eq|basic_sigma}) can be discretized on a time grid $t_i$ with $i=0,\ldots,n-1$ as follows
\begin{widetext}
\begin{equation}\label{eq|Euler}
\ln\sigma(t_{i+1})=\ln\sigma(t_{i})+\cfrac{1}{2}\,\cfrac{\sigma^2(t_{i})}{\delta_c^2(t_i)}\, \left|\cfrac{\delta_c(t_{i+1})-\delta_c(t_i)}{\delta_c(t_i)}\right|-\cfrac{\sigma(t_{i})}{\delta_c(t_i)}\, \left|\cfrac{\delta_c(t_{i+1})-\delta_c(t_i)}{\delta_c(t_i)}\right|^{1/2}\, w_i~,
\end{equation}
\end{widetext}
where $w_i$ are random weights extracted from a normal distribution with zero mean and unit variance. In Fig.~\ref{fig|sims} we show the resulting evolution of $\ln\sigma(M(t))$ as a function of cosmic time, with initial condition $M\sim 10^4\, M_\odot$ or $\ln\sigma\sim 2.2$ at $z\sim 10$ (reasonable initial conditions do not change significantly the outcome). The grey lines are $30$ randomly chosen evolutionary tracks, while the red line and shaded area illustrate the median and the quartiles over $3000$ realization of the noise, and the dashed blue line shows the evolution of the spherical collapse threshold $\delta_c(t)$. It is easily seen that the noise induces a drift of $\ln\sigma$, making it to decrease, and hence making the mass $M$ to increase. Remarkably, after a burn-in period needed to erase memory of the initial condition, $\sigma(M(t))$ tends to copy the evolution of $\delta_c(t)$; this means that the noise-induced drift effectively renders the standard hierarchical clustering of the halo population, as expressed by the increase with time of the characteristic mass $M_c(t)$ set by the condition $\sigma(M_c(t))\sim \delta_c(t)$.

\subsection{Fokker-Planck equation and the Press \& Schechter mass function}\label{sec|fokker}

We now look for the probability density $\mathcal{P}(M,t)$ for a region to be in a state of mass between $M$ and $M+{\rm d}M$ at time $t$; this can be found by solving the Fokker-Planck equation associated to Eq.~(\ref{eq|basic_mass}), which reads (see Appendix A for details)
\begin{widetext}
\begin{equation}\label{eq|fokker}
\cfrac{\partial}{\partial t} \mathcal{P}(M,t) = -T^2(t)\, \cfrac{\partial}{\partial M}\left[D(M)\, D'(M)\,\mathcal{P}(M,t)\right] +T^2(t)\,\cfrac{\partial^2}{\partial M^2} \left[D^2(M)\, \mathcal{P}(M,t)]\right]~,
\end{equation}
\end{widetext}
where we have defined the two quantities $D(M)\equiv \sigma^2/|{\rm d}\sigma/{\rm d}M|$ and $T(t)\equiv |\dot \delta_c|^{1/2}/\delta_c^{3/2}$ such that $\dot M = D(M)\, T(t)\, \eta(t)$ factorizes the mass and time dependencies. The Fokker-Planck equation may also be written as a pure continuity equation $\partial_t \mathcal{P}+\partial_M\, \mathcal{J}=0$ in terms of a probability current $\mathcal{J}(M,t)=-T^2(t)$ $ D(M)\, \partial_M\, [D(M)\, \mathcal{P}(M,t)]$.
The natural boundary conditions $\lim_{M\rightarrow \infty} \mathcal{P}(M,t)=0$, $\mathcal{P}(M,0)=\delta_{\rm D}(M)$ and the constraint $\mathcal{P}(M,t)=0$ whenever $M<0$ must apply; the latter corresponds to a reflective barrier condition $\mathcal{J}|_{M=0}=[D\, \partial_M\,(D\,\mathcal{P})]|_{M=0}=0$ at the $M=0$ point (no net probability current through $M=0$). Then the probability mass function $\mathcal{P}$ is normalized as $\int_0^\infty{\rm d}M\,
\mathcal{P}=1$ and thus it must be related to the halo mass function by
\begin{equation}\label{eq|PS}
\cfrac{{\rm d}N}{{\rm d}M\, {\rm d}V}(M,t) = \cfrac{\bar \rho_{\rm M}}{M}\, \mathcal{P}(M,t)~,
\end{equation}
in terms of the average comoving matter density $\bar\rho_{\rm M}$.

Now to solve the Fokker-Planck equation we employ the transformations:
\begin{equation}\label{eq|auxvar}
\left\{
\begin{aligned}
& X\equiv \int\cfrac{{\rm d}M}{D(M)} = \cfrac{1}{\sigma}\\
\\
& Y\equiv \int{\rm d}t\, T^2(t)=\cfrac{1}{2\, \delta_c^2}\\
\\
& \mathcal{W}(X,Y)\equiv D(M)\, \mathcal{P}(M,t)~.
\end{aligned}
\right.
\end{equation}
Then Eq.~(\ref{eq|fokker}) turns into
\begin{equation}\label{eq|diffeq}
\partial_Y\, \mathcal{W}(X,Y) = \partial^2_X\,\mathcal{W}(X,Y)~,
\end{equation}
which is a standard diffusion equation. In terms of these new variables, the boundary conditions stated below Eq.~(\ref{eq|fokker}) read $\lim_{X\rightarrow \infty}\mathcal{W}=0$, $\mathcal{W}(X,0)=\delta_{\rm D}(X)$ and $(\partial_X\,\mathcal{W})|_{X=0}=0$. These stem from the following circumstances: (i) for reasonable power spectra $\sigma$ is a slowly-varying inverse function of $M$, so that $X\propto \sigma^{-1}$ tends to zero or infinity as $M$ does; (ii) the collapse threshold $\delta_c\propto D^{-1}(t)$ scales inversely with $t$, so that
$Y\propto \delta_c^{-2}$ tends to zero as $t$ does; (iii) finally, $\mathcal{W}(X,Y) = D(M)\,\mathcal{P}(M,t)$ vanishes for large $X$ since $\mathcal{P}(M,t)$ is expected to be exponentially suppressed for $M\rightarrow \infty$ so overwhelming any slow (at most powerlaw) divergence of $D(M)=\sigma^2/|{\rm d}\sigma/{\rm d}M|$.

The solution of this differential problem is standard (it can be easily found via a Fourier transform) and writes
\begin{equation}\label{eq|diffsol}
\mathcal{W}(X,Y) = \cfrac{1}{\sqrt{\pi\, Y}}\, e^{-X^2/4\, Y}~.
\end{equation}
Coming back to the original variables we get
\begin{equation}\label{eq|fokkersol}
\mathcal{P}(M,t) = \cfrac{1}{D(M)\,\sqrt{\pi\, Y(t)}}\, e^{-X^2(M)/4\, Y(t)}~,
\end{equation}
which after Eqs. (\ref{eq|PS}) and (\ref{eq|auxvar}) yields the Press \& Schechter mass function:
\begin{empheq}[box=\fbox]{align}
N(M,t) = \sqrt{\cfrac{2}{\pi}} \,\cfrac{\bar \rho_{\rm M}\, \delta_c(t)}{M\,\sigma^2}
\left|\cfrac{{\rm d}\sigma}{{\rm d}M}\right|\, e^{-\delta_c^2(t)/2\, \sigma^2}~.
\end{empheq}

Three interesting remarks follow. First, note that the presence of the multiplicative noise in Eqs.~(\ref{eq|basic_sigma}) and (\ref{eq|basic_mass}) is fundamental in originating a second-order, diffusion-like term in the associated Fokker-Planck equation (see also Appendix A); from the derivation above it is seen that such a term yields the exponential cut-off of the mass function at the high-mass end. Second, one can easily compute the moments
\begin{equation}\label{eq|moments}
\left\langle\left(\cfrac{\delta_c}{\sigma}\right)^k\right\rangle = \int_0^{\infty}{\rm d}M\, (\delta_c/\sigma)^k\, \mathcal{P} = \cfrac{2^{k/2}}{\sqrt{\pi}}\, \Gamma\left(\cfrac{1+k}{2}\right)~.
\end{equation}
For scale-free power spectra $M\propto \sigma^{-6/(n+3)}$ holds in terms of the effective spectral index $n>-3$, so that the above implies $\langle M^k\rangle(t)\propto \delta_c^{-6\,k/(n+3)}$; this in turn scales as $\langle M^k\rangle(t)\propto t^{4\,k/(n+3)}$ in the redshift range where $\delta_c\propto t^{-2/3}$ applies.
Third, it is interesting to note that the Fokker-Planck Eq.~(\ref{eq|fokker}) highlights that the evolution of the probability density function is driven by the two terms on the r.h.s: the first represents its noise-induced drift toward larger masses because of hierarchical collapses, and the second describes its diffusive reshaping at the high-mass end due to the stochasticity in merging/accretion events.

\subsection{Stationarity}\label{sec|stationary}

An alternative derivation, that will be useful for the generalization in the next Section, is the following. We again start from Eq.~(\ref{eq|basic_sigma}) and change variable from $\sigma$ to $\nu\equiv \delta_c(t)/\sigma(M)$:
\begin{equation}\label{eq|basic_nu}
\cfrac{\rm d}{{\rm d}t}\,\nu = -\left|\cfrac{\dot\delta_c}{\delta_c}\right|\,\nu+ \left|\cfrac{\dot\delta_c}{\delta_c}\right|^{1/2}\, \eta(t)~;
\end{equation}
thus now the variable $\nu$ is seen to undergo a stochastic Ornstein-Uhlenbeck process. The corresponding Fokker-Planck equation reads
\begin{equation}\label{eq|fokkernu}
\partial_t\,\mathcal{P}(\nu,t)=\left|\cfrac{\dot\delta_c}{\delta_c}\right|\,
\partial_\nu\,\left[\nu\,\mathcal{P}(\nu,t)+\partial_\nu\,\mathcal{P}(\nu,t)\right]~.
\end{equation}
We set the boundary conditions $\lim_{\nu\rightarrow \infty}\mathcal{P}=0$ and $\mathcal{J}|_{\nu=0}=-|\dot\delta_c/\delta_c|\,[\nu\,\mathcal{P}+\partial_\nu\,\mathcal{P}]|_{\nu=0}=0$, implying $\int_0^\infty{\rm d}\nu\, \mathcal{P}=1$; the former is the natural boundary due to the expected exponential suppression of the mass function for large values of $M$, which correspond to large $\nu\propto \sigma^{-1}$; the latter is the no-current boundary condition at the $\nu=0$ point enforced by the constraint $\mathcal{P}(\nu,t)=0$ for $\nu<0$ being $\nu$ positively defined.

Having incorporated the time dependent quantity $\delta_c(t)$ into the new variable $\nu$, under the ergodic hypothesis one expects that the relevant solution in terms of this variable should be stationary (e.g., Paul \& Baschnagel 2013), i.e., it should satisfy $\partial_t\,\mathcal{P}(\nu,t)=0$.
From Eq.~(\ref{eq|fokkernu}), $\mathcal{P}(\nu,t)=\bar{\mathcal{P}}(\nu)$ is determined by
\begin{equation}\label{eq|currentnu}
\nu\,\bar{\mathcal{P}}+\cfrac{{\rm d}}{{\rm d}\nu}\,\bar{\mathcal{P}}=0~,
\end{equation}
where the constant on the r.h.s. must be zero to satisfy the no-current boundary condition $\mathcal{J}|_{\nu=0}=0$. The solution to this simple ordinary differential equation with normalization $\int_0^\infty{\rm d}\nu\, \bar{\mathcal{P}}(\nu)=1$ is $\bar{\mathcal{P}}=\sqrt{2/\pi}\, e^{-\nu^2/2}$. The mass function writes as
\begin{equation}\label{eq|massfuncnu}
N(M,t) = \cfrac{\bar \rho_{\rm M}}{M}\, \left|\cfrac{{\rm d}\nu}{{\rm d}M}\right|\, \bar{\mathcal{P}}(\nu)~;
\end{equation}
given that $|{\rm d}\nu/{\rm d}M| = (\delta_c/\sigma^2)\, |{\rm d}\sigma/{\rm d}M|$, this is indeed easily recognized to be again the Press \& Schechter mass function.

One may wonder whether the general time-dependent solution   $\mathcal{P}(\nu,t|\nu',t')$ of Eq.~(\ref{eq|fokkernu}) for a generic initial condition $\mathcal{P}(\nu,t=t'|\nu',t')=\delta_D(\nu-\nu')$ converges to the stationary state, and over which timescale. To this purpose, we note that defining a new time variable $\tau\propto -\ln\delta_c(t)$ brings Eq.~(\ref{eq|fokkernu}) into a form that can be easily solved via a Fourier transform; the fundamental solutions (a general result for Gaussian and Markovian variables known as Doob's theorem) read $\mathcal{P}\propto e^{-(\nu\pm\nu'\,e^{-\tau})^2/(1-e^{-2\tau})}/\sqrt{1-e^{-2\tau}}$. Taking into account the no-current boundary condition, and reverting to the original time variable one obtains
\begin{widetext}
\begin{equation}\label{eq|transnu}
\mathcal{P}(\nu,t|\nu',t') = \cfrac{1}{\sqrt{2\pi\,(1-\delta^2/\delta'^2)}}\,
\left\{\exp\left[{-\cfrac{1}{2}\,\cfrac{(\nu-\nu'\,\delta/\delta')^2}{1-\delta^2/\delta'^2}}\right]+
\exp\left[{-\cfrac{1}{2}\,\cfrac{(\nu+\nu'\,\delta/\delta')^2}{1-\delta^2/\delta'^2}}\right]\right\}~,
\end{equation}
\end{widetext}
with $\delta\equiv\delta_c(t)$ and $\delta'\equiv\delta_c(t')$. Plainly, away from any initial condition, for $\delta$ substantially lower than $\delta'$, this converges to the stationary solution $\mathcal{P}(\nu,t)$ derived above.
Such transitional states could be possibly related to deviation of the mass function from the self-similar shape Eq.~(\ref{eq|massfuncnu}) in terms of the variable $\nu$. Note, in passing, that the transition probability Eq.~(\ref{eq|transnu}) cannot be directly related to the halo conditional mass function of the extended Press \& Schechter theory; we anticipate that to derive the latter a modification of Eq.~(\ref{eq|basic_nu}) is needed, but the issue demands an extended analysis that will be presented in a forthcoming paper.

\section{Mass-dependent threshold: the $N$-body mass function}\label{sec|barrier}

\begin{figure}[!t]
\centering
\includegraphics[width=\columnwidth]{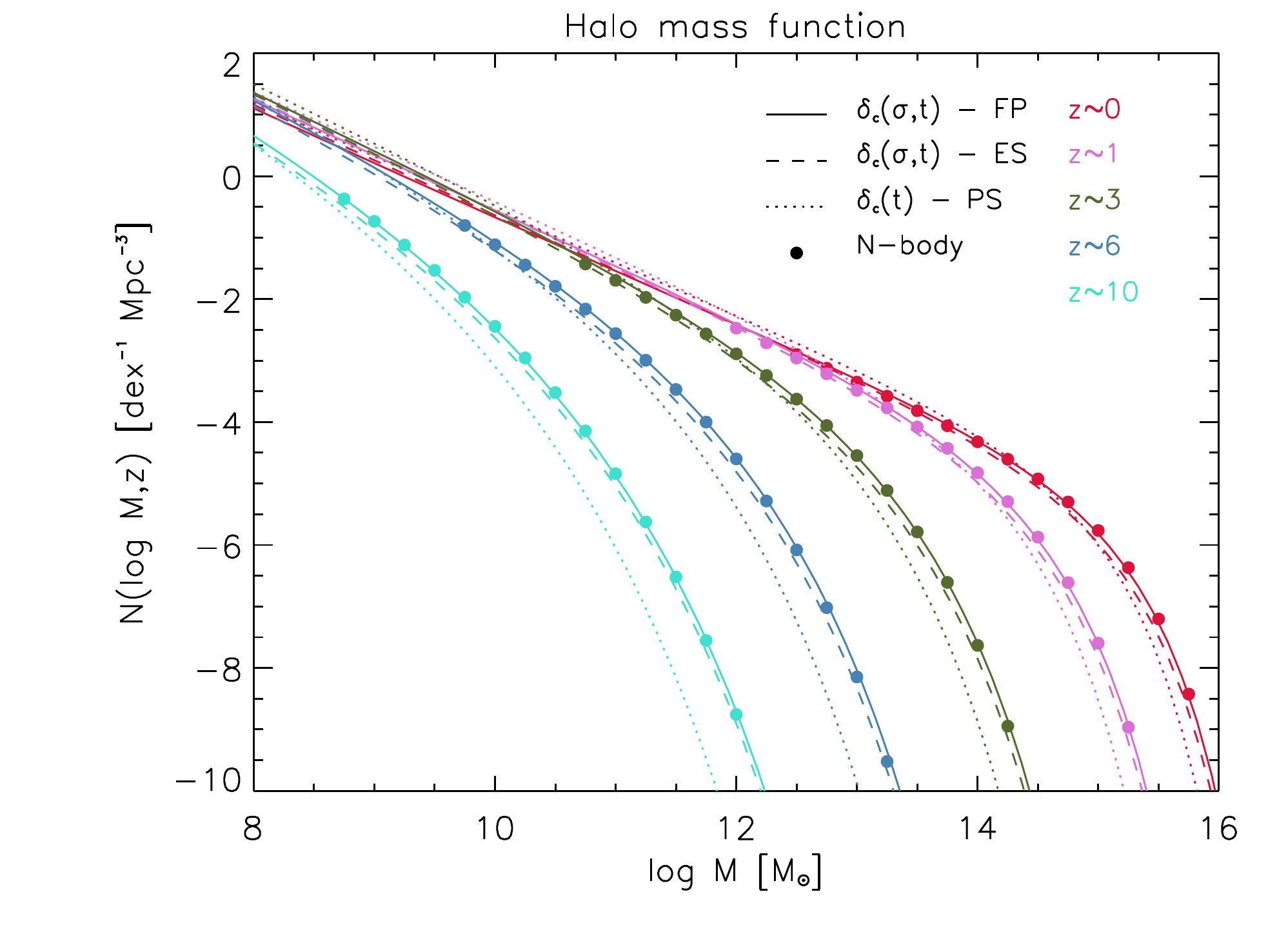}
\caption{Halo mass function at different redshifts $z=0$ (red), $1$ (magenta), $3$ (green), $6$ (blue), and $10$ (cyan). Solid lines refer to the mass function from this work based on the stationary Fokker-Planck solution (see Eqs.~\ref{eq|fokkersol_barrier_full} and \ref{eq|fokkersol_barrier_full2}) for a mass-dependent collapse threshold $\delta_c(\sigma,t)$, see Sect.~\ref{sec|barrier} for details. Filled dots illustrate the $N$-body results for FoF halos by Bhattacharya et al. (2011), sampled from their fitting formula for $-0.6\leq \ln[\sigma(M)\,D(z)]^{-1}\leq 1.3$ in mass bins of $0.25$ dex. Dashed line shows the mass function computed from the excursion set theory. As a reference, dotted lines show the Press \& Schechter mass function, which corresponds to the standard spherical collapse threshold $\delta_c(t)$ independent of mass.}\label{fig|massfunc}
\end{figure}

It is well known that the halo mass function derived from $N$-body simulations deviates substantially from the Press \& Schechter shape (see reference in Sect.~\ref{sec|intro} for details); in particular, the former is flatter than the latter both at the high and at the low mass end, and evolves more slowly toward high redshift (see Fig.~\ref{fig|massfunc}).  This mismatch is usually cured by introducing a mass-dependent threshold for collapse:
\begin{equation}\label{eq|barrier}
\delta_c(\sigma,t) = \sigma\, \sqrt{q}\, \nu\, \left[1+\cfrac{\beta}{(\sqrt{q}\, \nu)^{2\,\gamma}}\right]\equiv \sigma\, B(\nu)=\delta_c\, \cfrac{B(\nu)}{\nu}~,
\end{equation}
where $q$, $\beta$, and $\gamma$ are parameters to be set by fitting the $N$-body outcomes. Such a modified threshold is generally ascribed, though a bit naively, to the fact that perturbations undergo an ellipsoidal rather than a spherical collapse (Sheth \& Tormen 2002; see also discussion by Mo et al. 2010). Note, however, that the above shape is quite general in describing a variety of phenomena that can influence the collapse, like tidal torques and angular momentum, cosmological constant, dynamical friction (see Del Popolo 2017 and references therein). In the excursion set framework, the parameters $q\approx 0.707$, $\beta\approx 0.47$ and $\gamma\approx 0.615$ are required to fit the $N$-body mass function; however, there is some degeneracy, in that for example a square-root barrier with $q\approx 0.55$, $\beta\approx 0.5$ and $\gamma\approx 0.5$ fits simulations equally well. We stress that values of $\gamma>1/2$ are of some concern in the excursion set framework, because they imply that some walks $\delta(\sigma)$ do not cross the barrier at all, an occurrence thought to represent fragmentation. Notice that no general analytical expression exists for the excursion set mass function, apart from very particular barrier shapes (e.g., the constant, linear or square-root barriers; see for example Mahmood et al. 2006; Giocoli et al 2007; Lapi et al. 2013) and in general one must recur to numerical solutions (see Zhang \& Hui 2006).

\begin{deluxetable}{lcccccccccccc}
\tabletypesize{\scriptsize}
\tablewidth{0pt}\tablecaption{Solution parameters for a mass-dependent threshold}\tablehead{Mass function & & $q$ & & $\beta$ & & $\gamma$}
\startdata
ST99  & & $0.62$ & & $0.16$ & & $0.37$\\
Bh+11 & & $0.69$ & & $0.09$ & & $0.42$\\
Wa+13 & & $0.69$ & & $0.12$ & & $0.37$\\
\enddata
\tablecomments{Values of the parameters $q$, $\beta$ and $\gamma$ are set by fitting the Fokker-Planck solution Eqs.~(\ref{eq|fokkersol_barrier_full}) and (\ref{eq|fokkersol_barrier_full2}) to the halo mass function from the $N$-body simulations by Sheth \& Tormen (1999; ST99), Bhattacharya et al. (2011; Bh+11) and Watson et al. (2013; Wa+13), see Sect.~\ref{sec|barrier} for details.}
\end{deluxetable}

In the framework presented here, we aim to show that a minimal modification of the basic Eq.~(\ref{eq|basic_sigma}), which incorporates a mass-dependent collapse threshold with shape analogous to the above Eq.~(\ref{eq|barrier}), will yield a mass function in excellent agreement with $N$-body simulations. Moreover, we will provide an analytic expression valid for any triples of parameters $q$, $\beta$, and $\gamma<1/2$, that in the limit $\nu>>1$ (i.e., large masses and/or early times) matches the empirical formula introduced since Sheth \& Tormen (1999); this will establish a direct connection between the parameters describing the barrier and the shape of the halo mass function (which is absent in the excursion set approach). We start from Eq.~(\ref{eq|basic_sigma}) by replacing, in the multiplicative noise term, the standard spherical collapse threshold $\delta_c(t)$ with the mass-dependent $\delta_c(\sigma,t)$ given above; thus now the ratio $\sigma/\delta_c(\sigma,t)$ modulates the noise toward enforcing collapse. We also retain the term $|\dot \delta_c/\delta_c|^{1/2}\simeq |\dot D(t)/D(t)|^{1/2}$ involving the characteristic timescale for the linear growth of perturbations (see discussion in Sect.~\ref{sec|whitenoise}).
When formulated in terms of $\nu$, Eq.~(\ref{eq|basic_sigma}) modified in such a way writes as
\begin{equation}\label{eq|basic_barrier}
\cfrac{\rm d}{{\rm d}t}\,\nu = -\nu\,\left|\cfrac{\dot\delta_c}{\delta_c}\right| +\cfrac{\nu}{B(\nu)}\,\left|\cfrac{\dot\delta_c}{\delta_c}\right|^{1/2}\, \eta(t)~,
\end{equation}
where $\delta_c(t)$ is the standard threshold for spherical collapse.
The corresponding Fokker-Planck equation reads:
\begin{equation}\label{eq|fokker_barrier}
\partial_t\,\mathcal{P}(\nu,t)=\left|\cfrac{\dot\delta_c}{\delta_c}\right|\,
\partial_\nu\,\left\{\nu\,\mathcal{P}(\nu,t)+\cfrac{\nu}{B(\nu)}\,\partial_\nu\,
\left[\cfrac{\nu}{B(\nu)}\,\mathcal{P}(\nu,t)\right]\right\}~,
\end{equation}
with boundary conditions $\lim_{\nu\rightarrow \infty}\mathcal{P}=0$ and $\mathcal{J}|_{\nu=0}=-|\dot\delta_c/\delta_c|\,\,[\nu\,\mathcal{P}+(\nu/B)\,\partial_\nu\,(\nu\, \mathcal{P}/B)]|_{\nu=0}=0$, implying the constraint $\int_0^\infty{\rm d}\nu\, \mathcal{P}=1$.

In analogy to the procedure followed in Sect.~\ref{sec|stationary}, we look for stationary solutions $\mathcal{P}(\nu,t)=\bar{\mathcal{P}}(\nu)$ with $\partial_t\, \bar{\mathcal{P}}=0$; one obtains the equation
\begin{equation}
\cfrac{\nu}{B(\nu)}\,\cfrac{{\rm d}}{{\rm d}\nu}\,
\left[\cfrac{\nu}{B(\nu)}\,\bar{\mathcal{P}}\right] = -\nu\,\bar{\mathcal{P}}~,
\end{equation}
where the integration constant must be null to satisfy the no-current boundary condition $\mathcal{J}|_{\nu=0}=0$. The above equation can be easily solved by multiplying both sides by $1/B$ and recognizing that it becomes  separable for the function $(\nu/B)\,\bar{\mathcal{P}}$; we find
\begin{equation}\label{eq|fokkersol_barrier}
\bar{\mathcal{P}}(\nu)= \mathcal{A}\,\cfrac{B(\nu)}{\nu}\,\exp\left[-\int{\rm d}\nu\, \cfrac{B^2(\nu)}{\nu}\right]~,
\end{equation}
where the normalization constant $\mathcal{A}$ is determined by the condition $\int_0^\infty{\rm d}\nu\, \bar{\mathcal{P}}(\nu)=1$. We stress that this result holds for any mass-dependent collapse threshold that can be expressed as $\delta_c(\sigma,t)\equiv \delta_c(t)\, B(\nu)/\nu$ in terms of the scaled variable $\nu$.

Specializing now to the shape of $B(\nu)$ from Eq.~(\ref{eq|barrier}), we note that for $\nu\rightarrow 0$ the behavior $\nu\,\bar{\mathcal{P}}(\nu)\propto B(\nu)\propto \nu^{1-2\gamma}$ applies; to satisfy the normalization constraint $\int_0^\infty{\rm d}\nu\, \bar{\mathcal{P}}=1$ one must require $\gamma<1/2$. Performing explicitly the integration, we get the closed form expression
\begin{widetext}
\begin{empheq}[box=\widefbox]{align}\label{eq|fokkersol_barrier_full}
\bar{\mathcal{P}}(\nu)= \mathcal{A}\, \sqrt{q}\,\left[1+\cfrac{\beta}{(\sqrt{q}\,\nu)^{2\gamma}}\right]\, \exp\left\{-q\,\cfrac{\nu^2}{2}\,\left[1+\cfrac{2\,\beta}{1-\gamma}\,\cfrac{1}{(\sqrt{q}\,\nu)^{2\gamma}}
+\cfrac{\beta^2}{1-2\,\gamma}\,\cfrac{1}{(\sqrt{q}\,\nu)^{4\gamma}}\right]\right\}~.
\end{empheq}
\end{widetext}
The resulting mass function just writes
\begin{equation}\label{eq|fokkersol_barrier_full2}
N(M,t)= \cfrac{\bar\rho}{M\,\sigma}\, \left|\cfrac{{\rm d}\sigma}{{\rm d}M}\right|\,\nu\,\bar{\mathcal{P}}(\nu)~;
\end{equation}
incidentally, note that the multiplicity function $f(\ln\sigma)$ used in some literature works is just $f(\ln\sigma)\equiv \nu\,\bar{\mathcal{P}}(\nu)$.
We stress that the above is an exact expression, valid for any triple of values $q$, $\beta$, $\gamma<1/2$; the Press \& Schechter function is recovered for $q=1$ and $\beta=0$. Remarkably, the asymptotic behavior for $\nu>>1$, corresponding to large masses and/or early cosmic times, is seen to produce a shape akin to the empirical fit of $N$-body simulations adopted since Sheth \& Tormen (1999); however, for finite $\nu$ the terms in the exponential are important and must be taken into account.

We now use a Levenberg-Marquardt least-squares minimization routine to fit
the multiplicity function $\nu\,\bar{\mathcal{P}}(\nu)$ to the simulation outcomes for FoF halos by Sheth \& Tormen (1999), Bhattacharya et al. (2011) and Watson et al. (2013), sampled from their (somewhat different) fitting formulas for $-0.6\leq \ln[\sigma(M)\,D(z)]^{-1}\leq 1.3$ in mass bins of $0.25$ dex. For Sheth \& Tormen (1999), we find best fit parameters $q\approx 0.62$, $\beta\approx 0.16$ and $\gamma\approx 0.37$, yielding a value of the normalization constant $\mathcal{A}\approx 0.63$. For Bhattacharya et al. (2011), we obtain $q\approx 0.69$, $\beta\approx 0.09$ and $\gamma\approx 0.42$, yielding $\mathcal{A}\approx 0.64$. For Watson et al. (2013), we get $q\approx 0.69$, $\beta\approx 0.12$ and $\gamma\approx 0.37$, yielding $\mathcal{A}\approx 0.66$. These triples of values, reported for convenience in Table 1, are consistent within the uncertainties in the simulation results and in the fitting procedure.

\begin{figure}[!t]
\centering
\includegraphics[width=\columnwidth]{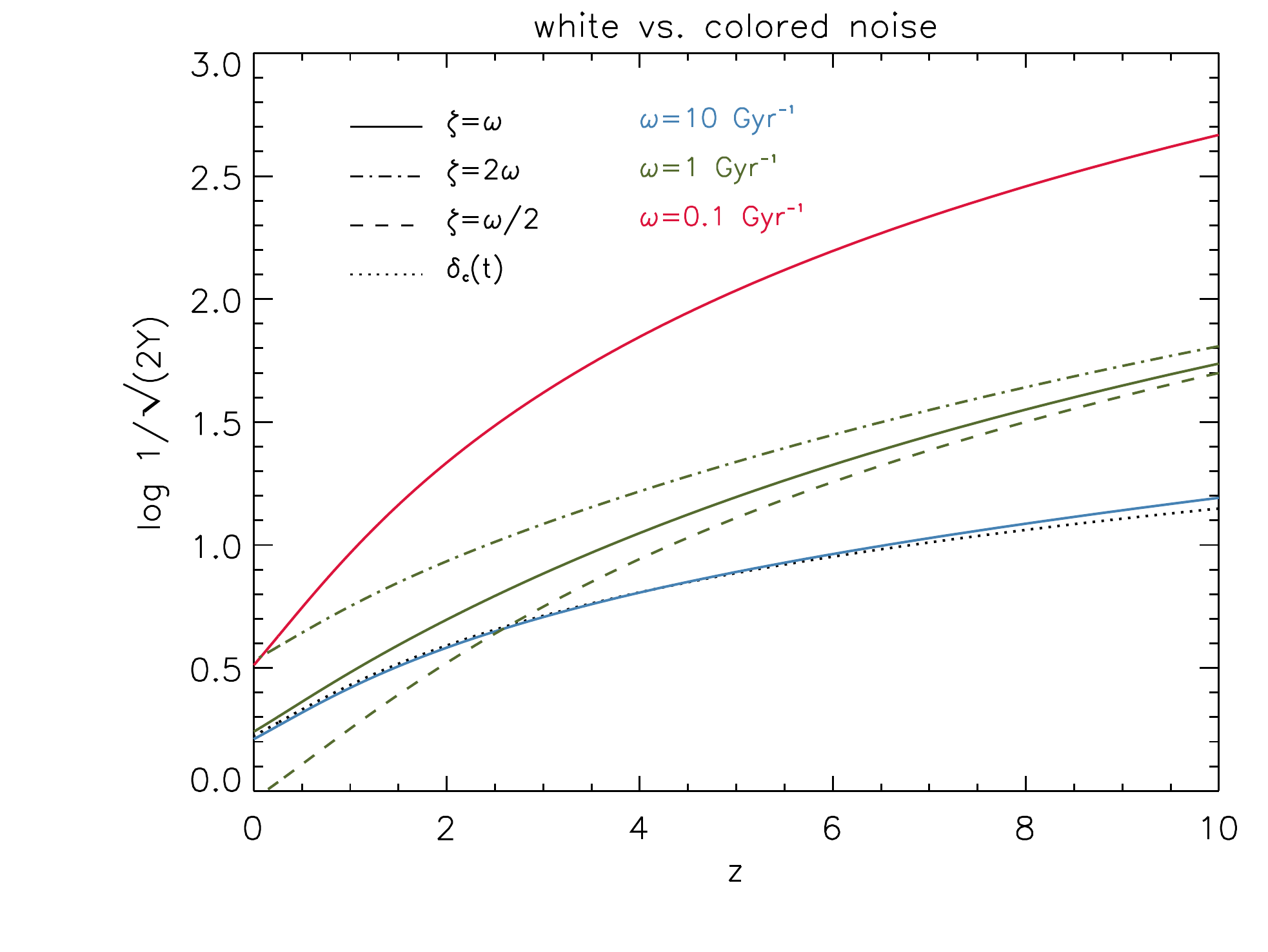}
\caption{Evolution with redshift of the quantity $(2\,Y)^{-1/2}$ entering the colored noise solution of Sect.~\ref{sec|colnoise}, that in the white-noise case is simply the spherical collapse threshold $\delta_c$ (dotted line). Colored curves illustrate the result for different values of the volatility $\zeta$ and mean reversal rate $\omega$ characterizing the noise: blue line is for $\omega=10$ Gyr$^{-1}$, green for $1$ Gyr$^{-1}$, and red for $0.1$ Gyr$^{-1}$; solid lines refer to $\zeta=\omega$, while dot-dashed line to $\zeta=2\,\omega$, and dashed line to $\zeta=\omega/2$ (these last two instances shown only for $\omega=1$ Gyr$^{-1}$).}\label{fig|colnoise}
\end{figure}

In Fig.~\ref{fig|massfunc} we compare the mass function $N(M,t)$ from Eqs.~(\ref{eq|fokkersol_barrier_full}) and (\ref{eq|fokkersol_barrier_full2})  to the $N$-body results by Bhattacharya et al. (2011), finding an excellent agreement over a wide range of masses $M\sim 10^8-10^{16}\, M_\odot$ and redshifts $z\sim 0-10$. In the same Figure we also plot for reference the Press \& Schechter mass function; moreover, we show the mass function computed from the excursion set approach (following the numerical algorithm by Zhang \& Hui 2006) with the barrier shape of Eq.~(\ref{eq|fokker_barrier}) and the standardly adopted parameters $q\approx 0.707$, $\beta\approx 0.47$ and $\gamma\approx 0.615$ (see above).

\section{Colored noise and non-Markovian walks}\label{sec|colnoise}

In Nature, white noise is never found to be perfectly realized, but rather constitutes an idealization of the stochastic driving force affecting a physical phenomenon. Thus one may wonder whether the previous treatment can be extended to a colored instead of a white noise; this will correspondingly enforce a non-Markovian evolution of the system. To have a grasp on the impact of colored noise and non-Markovianity on the mass function, we consider a multiplicative stochastic Ornstein-Uhlenbeck process; for simplicity we adopt a mass-independent collapse threshold, so that the endpoint of this computation should be compared with the Press \& Schechter mass function. Specifically, we modify our  Eq.~(\ref{eq|basic_mass}) into the two-dimensional stochastic system
\begin{equation}
\left\{
\begin{aligned}
& \dot M = T(t)\, D(M)\, \Gamma(t)\\
\\
& \dot\Gamma = -\omega\, \Gamma(t)+\zeta\, \eta(t)\\
\end{aligned}
\right.
\end{equation}
where, besides already defined quantities, $\Gamma(t)$ is an Ornstein-Uhlenbeck noise with
average $\langle\Gamma(t)\rangle=0$ and a nontrivial correlation between different times
\begin{equation}
\langle\Gamma(t)\,\Gamma(t')\rangle = \cfrac{\zeta^2}{\omega}\,[ e^{-\omega\,|t-t'|}-e^{-\omega(t+t')}]~,
\end{equation}
controlled by the parameters $\omega$ and $\zeta$ (both have physical dimension of $1/$time); $\zeta$ represents the degree of volatility, i.e. the sensitivity of the system to random changes, while $\omega$ is the dissipation rate at which the system tends to reverse toward its zero mean. This is perhaps the simplest generalization of the white noise case, since $\Gamma(t)$ is still Markovian, but of course $M(t)$ is not. In the limit $\zeta=\omega\rightarrow \infty$ one recovers a pure white-noise, since $\langle\Gamma(t)\,\Gamma(t')\rangle \simeq \lim_{\omega\rightarrow \infty}\, \omega\,e^{-\omega\,|t-t'|} = 2\, \delta_{\rm D}(t-t')$.

The Fokker-Planck equation regulating the dynamics of the probability density function $\mathcal{P}(M,\Gamma,t)$ for the above system is
\begin{widetext}
\begin{equation}\label{eq|coloredfokker}
\cfrac{\partial}{\partial t}\mathcal{P}(M,\Gamma,t) = -T(t)\, \Gamma\,\cfrac{\partial}{\partial M}\left[D(M)\, \mathcal{P}(M,\Gamma,t)\right]+\omega\, \cfrac{\partial}{\partial \Gamma}\left[\Gamma\, \mathcal{P}(M,\Gamma,t)\right]+\zeta^2\,  \cfrac{\partial^2}{\partial \Gamma^2}\,\mathcal{P}(M,\Gamma,t)~;
\end{equation}
\end{widetext}
in analogy with the one-dimensional case, we can write this is as a continuity equation $\partial_t \mathcal{P}+\nabla\cdot\mathcal{J}=0$ in terms of the vectorial differential operator $\nabla=[\partial_M,\partial_\Gamma]$ and probability current $\mathcal{J}=[\mathcal{J}_M,\mathcal{J}_\Gamma]=[T\, D\, \Gamma\,\mathcal{P},-\omega\Gamma\,\mathcal{P}-\zeta^2\,\partial_\Gamma\, \mathcal{P}]$. In particular, we are interested in the marginalized $\mathcal{P}(M,t)=\int{\rm d}\Gamma\, \mathcal{P}(M,\Gamma,t)$ with boundary conditions $\lim_{M\rightarrow \infty}\,\mathcal{P}(M,t)=0$, $\mathcal{P}(M,0)=\delta_{\rm D}(M)$ and $(\int{\rm d}\Gamma\, \Gamma\, \mathcal{P})|_{M=0}=0$; the latter expresses in the two-dimensional space $(M,\Gamma)$ the requirement of a zero current on the $M=0$ line, i.e. $(\int{\rm d}\Gamma\, \mathcal{J}_M)|_{M=0}=0$. In Appendix B we show that the Fokker-Planck equation is solved by
\begin{equation}
\mathcal{P}(M,t) = \cfrac{1}{D\,\sqrt{\pi\, Y}}\, e^{-X^2/4\,Y}~,
\end{equation}
in terms of:
\begin{equation}
\left\{
\begin{aligned}
&X(M)=\int\cfrac{{\rm d}M}{D(M)}=\cfrac{1}{\sigma(M)} \\
\\
&Y(t)=\cfrac{\zeta^2}{\omega}\,\int^t{\rm d}\tau\, T(\tau)\, e^{-\omega\, \tau}\, \int^\tau{\rm d}\tau'\, T(\tau')\,[e^{\omega\,\tau'}-e^{-\omega\,\tau'}]~,\\
\end{aligned}
\right.
\end{equation}
This is the equivalent for colored noise of Eq.~(\ref{eq|fokkersol}), which is recovered in the limit $\zeta=\omega\rightarrow \infty$ as
$Y(t)\rightarrow \int{\rm d}t\, T^2(t)=1/2\,\delta_c^2$.
The corresponding expression for the mass function is written as:
\begin{equation}
N(M,t)= \cfrac{1}{\sqrt{\pi\, Y(t)}}  \,\cfrac{\bar \rho_{\rm M}}{M\,\sigma^2}\,
\left|\cfrac{{\rm d}\sigma}{{\rm d}M}\right|\, e^{-1/4\,\sigma^2\,Y(t)}~;
\end{equation}
this is similar to the Press \& Schechter shape, but for a modified redshift evolution encoded in $Y(t)$. The quantity $(2\, Y)^{-1/2}$, that in the white-noise limit is just $\delta_c$, can be regarded as a modified collapse threshold; in Fig.~\ref{fig|colnoise} we show how
its evolution and absolute value differ from $\delta_c(t)$, depending on the volatility $\zeta$ and mean-reversal rate $\omega$ characterizing the noise.
We conclude that only values of $\zeta\sim\omega\gtrsim$ several Gyr$^{-1}$
are required not to move far away from the Press \& Schechter mass function, and hence from simulations.

\section{Summary and outlook}\label{sec|summary}

In this paper we have submitted a new theory of the hierarchical clustering based on stochastic differential equations in real space, that constitutes a change of perspective with respect to the excursion set formalism; this work is specifically focused on the halo mass function.

First, we have invented a stochastic differential equation that describes fluctuations in the mass growth of DM halos, as driven by a multiplicative white (Gaussian) noise dependent on the spherical collapse threshold and on the power spectrum of DM perturbations. By numerically integrating such a stochastic differential equation, in Sect.~\ref{sec|whitenoise} we have demonstrated that the noise yields an average drift of the halo population toward larger masses, that quantitatively renders the standard hierarchical clustering (see Fig.~\ref{fig|sims}). Then, in Sect.~\ref{sec|fokker} we have solved the Fokker-Planck equation associated to the stochastic dynamics, and obtained as a solution the Press \& Schechter mass function;  in Sect.~\ref{sec|stationary} we have pointed out that the solution is stationary when the original equation is written in convenient variables.

Then in Sect.~\ref{sec|barrier} we have introduced a minimal modification of the stochastic equation in terms of a mass-dependent collapse threshold. Using a parametric shape analogous to that adopted in the excursion set framework, we have obtained a closed-form analytical solution of the associated Fokker-Planck equation. Remarkably, such a solution has a limiting shape for large masses/early times similar to the empirical fitting formula introduced since Sheth \& Tormen (1999); in fact, for specific values of the parameters describing the mass dependence of the collapse threshold, our result reproduces extremely well the $N$-body mass function over a wide range of masses and redshifts (see Fig.~\ref{fig|massfunc}).

As an aside issue, in Sect.~\ref{sec|colnoise} we have generalized our stochastic approach to a colored, instead of a white, noise; in particular, we have investigated the modification to the Press \& Schechter mass function when the stochastic dynamics is ruled by a multiplicative Ornstein-Uhlenbeck noise with finite volatility and mean-reversal rate. We have exactly solved the related Fokker-Planck equation, finding that the mass function has shape analogous to the Press \& Schechter one when expressed in terms of a modified, effective collapse threshold; the latter may substantially differ from the standard $\delta_c(t)$ in absolute value and time evolution, depending on the correlation parameters of the noise (see Fig.~\ref{fig|colnoise}). We conclude that values of such parameters larger than several Gyr$^{-1}$ are required not to move far away from the Press \& Schechter mass function, and hence from simulations.

The next-order development of this work will concern the computation of the conditional mass function, i.e., the mass function of a halo's progenitors. This investigation will naturally extend to merger rates, formation time distributions, and large-scale halo bias. A more detailed comparison of our results with the outcomes of $N$-body simulations, that includes the specificity of both the numerical experiments as well as of the theory, will be welcome. Other future applications could involve a re-examination of the two-phase mass growth of DM halos, the halo specific angular momentum distribution, the void mass function, and halo statistics in non-standard cosmological frameworks. We very much hope that the new perspective offered by the theory presented here will contribute to a better understanding of the gravitational dynamics leading to the formation and evolution of DM halos and hosted baryonic structures across cosmic times.

\begin{acknowledgements}
We thank our referee for a constructive report, and for the insightful comments and helpful suggestions. We acknowledge Carlo Baccigalupi, Alessandro Bressan, and Giovanni Bussi for enlightening discussions and critical reading. This work has been partially supported by PRIN MIUR 2017 prot. 20173ML3WW 002, `Opening the ALMA window on the cosmic evolution of gas, stars and supermassive black holes'. A.L. has taken advantage of the MIUR grant `Finanziamento annuale individuale attivit\'a base di ricerca' and of the EU H2020-MSCA-ITN-2019 Project 860744
`BiD4BEST: Big Data applications for Black hole Evolution STudies'.
\end{acknowledgements}

\begin{appendix}

\section{A. A primer on the stochastic differential and Fokker-Planck equations}

Given that concepts and techniques related to the stochastic differential and Fokker-Planck equations are not very common among the astrophysics community, for the reader's convenience we present here a short primer, in a modern notation and systematic way. In particular, we focus on the derivation of the Fokker-Planck equation associated to a given stochastic system, in presence of a state-dependent, multiplicative noise; this is extensively used in the main text. More details and applications can be found, e.g., in the book by Risken (1996).

The derivation involves two steps: (i) an expression for the time derivative of the probability density in terms of a Taylor-series of the conditional moments, known as Kramers-Moyal expansion; (ii) the explicit computation of such moments for a random variable satisfying a stochastic differential equation with multiplicative noise. Suppose we are given a system characterized by a physical variable $\xi$, whose evolution $\xi(t)$ as a function of time $t$ is stochastic. If the evolution is Markovian, by definition the probability density function $\mathcal{P}(x,t)$ of finding the system in state $\xi(t)=x$ at time $t$ satisfies:
\begin{equation}\label{eq|transitionprob}
\mathcal{P}(x,t+\tau)=\int{\rm d}x'\, \mathcal{P}(x,t+\tau|x',t)\, \mathcal{P}(x',t)~,
\end{equation}
in terms of the conditional (transition) probability $\mathcal{P}(x,t+\tau|x',t)$ between the times $t$ and $t+\tau$; in other words, for a Markovian system the transition probability depends only on the value at the next earlier time. We rewrite the integrand as $\mathcal{P}(x,t+\tau|x',t)\, \mathcal{P}(x',t)=\mathcal{P}(x+\Delta-\Delta,t+\tau|x-\Delta,t)\, \mathcal{P}(x-\Delta,t)$ in terms of $\Delta\equiv x-x'$ and then perform a Taylor expansion in $\Delta$ to obtain
\begin{equation}
\mathcal{P}(x,t+\tau|x',t)\, \mathcal{P}(x',t) \simeq \sum_{n=0}^{\infty}\, \cfrac{(-1)^n}{n!}\, \Delta^n\, \partial_x^n\, [\mathcal{P}(x+\Delta,t+\tau|x,t)\, \mathcal{P}(x,t)]~.
\end{equation}
Now we insert this expression in Eq.~(\ref{eq|transitionprob}) and perform the integration after changing variable from $x'$ to $\Delta$; noting that in the $n=0$ term $\int{\rm d}\Delta\, \mathcal{P}(x+\Delta,t+\tau|x,t)=1$ holds since the conditional probability is normalized, we get
\begin{equation}
\mathcal{P}(x,t+\tau)-\mathcal{P}(x,t)\simeq \sum_{n=1}^{\infty}\, \cfrac{1}{n!}\,(-\partial_x)^n\, \mathcal{M}_n(x,t;\tau)~,
\end{equation}
where we have defined the conditional moments
\begin{equation}\label{eq|condmoments}
\mathcal{M}_n(x,t;\tau) \equiv \langle|\xi(t+\tau)-\xi(t)|^n\rangle|_{\xi(t)=x} = \int{\rm d}\Delta\,\Delta^n\, \mathcal{P}(x+\Delta,t+\tau|x,t)~.
\end{equation}
Now we Taylor-expand the moments with respect to $\tau$ as follows
\begin{equation}
\mathcal{M}_n(x,t;\tau)/n! \simeq \mathcal{D}_n(x,t)\, \tau + \vartheta(\tau^2)~;
\end{equation}
note that terms of order $\tau^0$ cannot be present since $P(x+\Delta,t|x,t)=\delta_{\rm D}(\Delta)$ by definition and in Eq.~(\ref{eq|condmoments}) all the conditional moments for $n\geq 1$ vanish. For future reference the coefficients $\mathcal{D}_n$ are defined as
\begin{equation}
\mathcal{D}_n(x,t)\equiv \lim_{\tau\rightarrow 0} \cfrac{1}{n!}\,\cfrac{\mathcal{M}_n(x,t;\tau)}{\tau} = \lim_{\tau\rightarrow 0} \cfrac{1}{n!}\,\cfrac{\langle|\xi(t+\tau)-\xi(t)|^n\rangle|_{\xi(t)=x}}{\tau}~.
\end{equation}
All in all, we obtain the so called Kramers-Moyal expansion in terms of the partial differential equation
\begin{equation} \label{eq|KramersMoyal}
\partial_t \mathcal{P}(x,t) \simeq \sum_{n=1}^{\infty}\, (-\partial_x)^n\, \mathcal{D}_n(x,t)~;
\end{equation}
this ends the first step in the derivation.

We now compute explicitly the coefficients $\mathcal{D}_n$ when the variable $\xi(t)$ satisfies a stochastic differential equation
\begin{equation}
\dot\xi = h(\xi,t)+g(\xi,t)\, \eta(t)~,
\end{equation}
with inital condition $\xi(t)=x$. Here $\eta(t)$ is a white (Gaussian) noise with ensemble-average properties $\langle\eta(t)\rangle=0$ and $\langle\eta(t)\, \eta(t')\rangle=2\, \delta_{\rm D}(t-t')$; the coefficient $2$ in this last expression is only a convenient arbitrary choice, since it can be reabsorbed into the multiplicative function $g$ without loss of generality. We start by transforming the differential into an integral stochastic equation
\begin{equation}
\xi(t+\tau)-x = \int_t^{t+\tau}{\rm d}t'\, [h(\xi(t'),t')+g(\xi(t'),t')\, \eta(t')]~.
\end{equation}
We expand near $x$ the functions $h(\xi(t'),t')\simeq h(x,t')+\partial_x h(x,t')\,(\xi(t')-x)+...$ and $g(\xi(t'),t')\simeq g(x,t')+\partial_x g(x,t')\,(\xi(t')-x)+...$ to obtain
\begin{equation}
\begin{aligned}
\xi(t+\tau)-x &\simeq \int_t^{t+\tau}{\rm d}t'\, h(x,t')+\int_t^{t+\tau}{\rm d}t'\, \partial_x h(x,t')\,(\xi(t')-x)+\ldots \\
\\
&+\int_t^{t+\tau}{\rm d}t'\, g(x,t')\, \eta(t')+\int_t^{t+\tau}{\rm d}t'\, \partial_x g(x,t')\, \eta(t')\,(\xi(t')-x)+\ldots\\
\end{aligned}
\end{equation}
Then we iterate for $\xi(t')-x$ in the integrand to get
\begin{equation}
\begin{aligned}
\xi(t+\tau)-x &= \int_t^{t+\tau}{\rm d}t'\, h(x,t')+\int_t^{t+\tau}{\rm d}t'\,\partial_x h(x,t')\,\int_t^{t'}{\rm d}t''\,h(x,t'') + \\
\\
& +\int_t^{t+\tau}{\rm d}t'\, \partial_x h(x,t')\,\int_t^{t'}{\rm d}t''\,g(x,t'')\,\eta(t'')+\ldots \\
\\
& +\int_t^{t+\tau}{\rm d}t'\, g(x,t')\, \eta(t')+
\int_t^{t+\tau}{\rm d}t'\,g(x,t')\,\int_t^{t'}{\rm d}t''\,h(x,t'')\, \eta(t'')+\\
\\
&+\int_t^{t+\tau}{\rm d}t'\,\partial_x g(x,t')\,\int_t^{t'}{\rm d}t''\,g(x,t'')\, \eta(t')\,\eta(t'') + \ldots\\
\end{aligned}
\end{equation}
Now taking the ensemble average and using the properties of the white noise yields
\begin{equation}
\begin{aligned}
\langle \xi(t+\tau)-x\rangle &= \int_t^{t+\tau}{\rm d}t'\, h(x,t')+\int_t^{t+\tau}{\rm d}t'\,\int_t^{t'}{\rm d}t''\,h(x,t'')\,\partial_x\,h(x,t')+\ldots\\
\\
&+\int_t^{t+\tau}{\rm d}t'\,g(x,t')\,\partial_x g(x,t')+\ldots\\
\end{aligned}
\end{equation}
where in the last term we have used that $\int_t^{t'}{\rm d}t''\,2\,\delta_{\rm D}(t''-t')\, g(x,t'')=g(x,t')$ since the Dirac-$\delta$ operates on an extremal of the integration.
Dividing by $\tau$ and taking the limit for $\tau\rightarrow 0$ one immediately recognizes the coefficient
$\mathcal{D}_1 = h(x,t)+g(x,t)\,\partial_x g(x,t)$.
For higher-order coefficients notice that terms containing the noise are proportional to $\tau^n$ where $n$ is the number of integrals involved, and vanish for small $\tau$; actually only one of such terms, containing two integrals and two noises contributes and yields
$\mathcal{D}_2 = \lim_{\tau\rightarrow 0}\,(1/2\tau)\, \int_t^{t+\tau}{\rm d}t'\,\int_t^{t'}{\rm d}t''\, g(x,t')$ $g(x,t'')\,2\,\delta_{\rm D}(t'-t'')=g^2(x,t)$, while $\mathcal{D}_n=0$ for any $n\geq 3$. This ends the second step of the derivation.

Putting together the coefficients just derived in the Kramers-Moyal expansion of Eq.~(\ref{eq|KramersMoyal}), one finds the Fokker-Planck equation corresponding to the original stochastic equation:
\begin{equation}
\left\{
\begin{aligned}
\partial_t \mathcal{P}(x,t) &=-\partial_x\, [\mathcal{D}_1(x,t)\,\mathcal{P}(x,t)]+\partial_x^2\, [\mathcal{D}_2(x,t)\,\mathcal{P}(x,t)]\\
\\
\mathcal{D}_1(x,t) & =h(x,t)+g(x,t)\,\partial_x\, g(x,t)\\
\\
\mathcal{D}_2(x,t) & =g^2(x,t)\\
\end{aligned}
\right.
\end{equation}
The quantity $g\, \partial_x\, g$ appearing in the coefficient $\mathcal{D}_1$ is a noise-induced drift; this stems from the fact that as $\eta(t)$ fluctuates, also the random variable $\xi(t)$ and so the function $g(\xi(t),t)$ varies and therefore $\langle g(\xi(t),t)\, \eta(t)\rangle$ is not null even if $\langle\eta(t)\rangle=0$ is. Finally, simple algebra shows that the Fokker-Planck equation may be written as a source-free continuity equation:
\begin{equation}\label{eq|continuity}
\partial_t \mathcal{P}(x,t)+\partial_x \mathcal{J}(x,t)=0
\end{equation}
in terms of a probability current
\begin{equation}\label{eq|current}
\mathcal{J}(x,t)\equiv h(x,t)\, \mathcal{P}(x,t)-g(x,t)\, \partial_x [g(x,t)\,\mathcal{P}(x,t)]~.
\end{equation}

\section{B. Solution of the Fokker-Planck equation for colored noise}

In this Appendix we show how to solve the Fokker-Planck equation derived in Sect.~\ref{sec|colnoise}
\begin{equation}
\partial_t \mathcal{P}(M,\Gamma,t) = -T(t)\, \Gamma\, \partial_M\,[ {D\,\mathcal{P}(M,\Gamma,t)}]+\omega\, \partial_\Gamma\,[\Gamma\, \mathcal{P}(M,\Gamma,t)]+\zeta^2\, \partial^2_\Gamma\, \mathcal{P}(M,\Gamma,t)~.
\end{equation}
for the marginalized $\mathcal{P}(M,t)=\int{\rm d}\Gamma\, \mathcal{P}(M,\Gamma,t)$ with boundary conditions $\lim_{M\rightarrow \infty}\, \mathcal{P}=0$, $\mathcal{P}(M,0)=\delta_{\rm D}(M)$ and $(\int{\rm d}\Gamma\, \Gamma\, \mathcal{P})|_{M=0}=0$.

As a first step, we introduce a new variable $X\equiv \int{\rm d}M/D(M)$ in place of $M$, and redefine the probability density as $\mathcal{W}(X,\Gamma,t)=D(M)\,\mathcal{P}(M,\Gamma,t)$; then the above equation turns into
\begin{equation}
\partial_t \mathcal{W} = -T(t)\, \Gamma\, \partial_X\,\mathcal{W}+\omega\, \partial_\Gamma\,[\Gamma\, \mathcal{W}]+\zeta^2\, \partial^2_\Gamma\, \mathcal{W}~.
\end{equation}
We now perform a two-dimensional Fourier transform
\begin{equation}\label{eq|fourtrans}
\mathcal{W}(X,\Gamma,t)\propto \int{\rm d}k_X\, {\rm d}k_\Gamma\, \tilde{\mathcal{W}}(k_X,k_\Gamma,t)\,e^{i\,(k_X\, X+k_\Gamma\, \Gamma)}~,
\end{equation}
and obtain the following equation for the Fourier modes
\begin{equation}
\partial_t \tilde{\mathcal{W}} = T(t)\, k_X\, \partial_{k_\Gamma}\, \tilde{\mathcal{W}}-\omega\, k_\Gamma\,\partial_{k_\Gamma} \tilde{\mathcal{W}}-\zeta^2\, k_\Gamma^2\, \tilde{\mathcal{W}}~.
\end{equation}
Given the boundary conditions, it is convenient to look for solutions with shape
\begin{equation}
\tilde{\mathcal{W}}(k_X,k_\Gamma,t) \propto e^{-k_X^2\, \Sigma_{XX}/2-k_\Gamma^2\, \Sigma_{\Gamma\Gamma}/2-k_X\,k_\Gamma\, \Sigma_{X\Gamma}}
\end{equation}
where $\Sigma_{XX}(t)$, $\Sigma_{X\Sigma}(t)$, $\Sigma_{\Sigma\Sigma}(t)$ are only functions of time. Inserting this ansantz into the previous equation yields the following ordinary differential equations
\begin{equation}
\left\{
\begin{aligned}
&\dot\Sigma_{\Gamma\Gamma}=-2\,\omega\,\Sigma_{\Gamma\Gamma}+2\, \zeta^2\\
\\
&\dot\Sigma_{X\Gamma}=T\,\Sigma_{\Gamma\Gamma}-\omega\,\Sigma_{X\Gamma}\\
\\
&\dot\Sigma_{XX}=2\, T\, \Sigma_{X\Gamma}~.\\
\end{aligned}
\right.
\end{equation}
These can be straightforwardly solved as
\begin{equation}
\left\{
\begin{aligned}
&\Sigma_{\Gamma\Gamma}(t)=\cfrac{\zeta^2}{\omega}\,(1-e^{-2\,\omega\,t})\\
\\
&\Sigma_{X\Gamma}(t)=\int^t{\rm d}\tau\, T(\tau)\,e^{-\omega\, (t-\tau)}\,\Sigma_{\Gamma\Gamma}(\tau)\\
\\
&\Sigma_{XX}(t)=2\, \int^t{\rm d}\tau\, T(\tau)\, \Sigma_{X\Gamma}(\tau)~.\\
\end{aligned}
\right.
\end{equation}
Inverting the Fourier transform in Eq.~(\ref{eq|fourtrans}) one finds out the solution
\begin{equation}
\mathcal{W}(X,\Gamma,t) \propto \cfrac{1}{\sqrt{||\Sigma||}}\, \exp\left\{-\cfrac{X^2}{2\,\Sigma_{XX}}-\cfrac{[\Sigma_{XX}\,\Gamma-\Sigma_{X\Gamma}\,X]^2}{2\, \Sigma_{XX}\, ||\Sigma||}\right\}~,\\
\end{equation}
where $||\Sigma||=\Sigma_{XX}\Sigma_{\Gamma\Gamma}-\Sigma_{X\Gamma}^2$ is the determinant of the $2\times 2$ symmetric matrix constructed with the $\Sigma$s; this can be easily checked to satisfy the desired boundary conditions, that in terms of the variable $X$ and function $\mathcal{W}$ read $\lim_{X\rightarrow \infty}\, \mathcal{W}=0$, $\mathcal{W}(X,0)=\delta_{\rm D}(X)$, and $(\int{\rm d}\Gamma\, \Gamma\, \mathcal{W})|_{X=0}=0$.

Finally, marginalizing over $\Gamma$ and coming back to the original variables, one obtains
\begin{equation}
\mathcal{P}(M,t) = \cfrac{1}{D\,\sqrt{\pi\, Y}}\, e^{-X^2/4\,Y}~,
\end{equation}
in terms of the quantities:
\begin{equation}
\left\{
\begin{aligned}
&X(M)=\int\cfrac{{\rm d}M}{D(M)} \\
\\
&Y(t)\equiv \cfrac{\Sigma_{XX}}{2} = \cfrac{\zeta^2}{\omega}\,\int^t{\rm d}\tau\, T(\tau)\, e^{-\omega\, \tau}\, \int^\tau{\rm d}\tau'\, T(\tau')\,[e^{\omega\,\tau'}-e^{-\omega\,\tau'}]\\
\end{aligned}
\right.
\end{equation}
note that the correct white-noise limit is recovered for $\zeta=\omega\rightarrow \infty$ since $Y(t)\rightarrow \int{\rm d}t\, T^2(t)=1/2\,\delta_c^2(t)$, and that actually $(2\,Y)^{-1/2}$ constitute an effective collapse threshold, dependent on the parameters of the colored noise.

\end{appendix}

\end{document}